\documentclass[aps, prd, amsmath, floats, floatfix, twocolumn, nofootinbib, superscriptaddress]{revtex4-1}
\usepackage[pdftex]{graphicx}
\usepackage{color}
\usepackage{latexsym}
\usepackage[colorlinks]{hyperref}
\usepackage{amsmath}

\newcommand{\be}{\begin{eqnarray}}
\newcommand{\ee}{\end{eqnarray}}

\begin{document}

\title{Implementation of a radial disk ionization profile in the {\tt relxill\_nk} model}

\author{Askar~B.~Abdikamalov}
\affiliation{Center for Field Theory and Particle Physics and Department of Physics, Fudan University, 200438 Shanghai, China}
\affiliation{Ulugh Beg Astronomical Institute, Tashkent 100052, Uzbekistan}

\author{Dimitry~Ayzenberg}
\affiliation{Theoretical Astrophysics, Eberhard-Karls Universit\"at T\"ubingen, D-72076 T\"ubingen, Germany}

\author{Cosimo~Bambi}
\email[Corresponding author: ]{bambi@fudan.edu.cn}
\affiliation{Center for Field Theory and Particle Physics and Department of Physics, Fudan University, 200438 Shanghai, China}

\author{Honghui~Liu}
\affiliation{Center for Field Theory and Particle Physics and Department of Physics, Fudan University, 200438 Shanghai, China}

\author{Yuexin~Zhang}
\affiliation{Kapteyn Astronomical Institute, University of Groningen, 9747 AD Groningen, The Netherlands}

\begin{abstract}
Very steep reflection emissivity profiles in the inner part of accretion disks are commonly found in the analysis of X-ray observations of black hole binaries and AGN, but there is some debate about their exact origin. While steep reflection emissivity profiles can be naturally produced by compact coronae close to black holes, the measured radial emissivity parameter can be further increased by the radial disk ionization profile when the theoretical model assumes a disk with constant ionization. In this paper, we implement the possibility of a radial disk ionization profile in the reflection model {\tt relxill\_nk}, which is a package designed to calculate reflection spectra of ``deformed'' Kerr black holes. We analyze a \textsl{NuSTAR} observation of the black hole binary EXO~1846--031, which was previously found to have a very high inner emissivity index. We find that the model with a radial disk ionization profile improves the fit, but the impact on the estimate of the black hole spin parameter and on the constraint of the deformation parameter is modest. However, we show that the analysis of future observations of \textsl{Athena} and \textsl{eXTP} will necessarily require models with a radial disk ionization profile to have accurate constraints of the deformation parameters.  
\end{abstract}

\maketitle


\section{Introduction}

Relativistic reflection features are common in the X-ray spectra of accreting black holes~\cite{Fabian:1989ej,Tanaka:1995en,Nandra:1996vv,Walton:2012aw,Tripathi:2020yts}. They are thought to be generated by illumination of the accretion disk by a hot corona, which is some energetic plasma near the black hole~\cite{Fabian:1995qz,Risaliti:2013cga}; for a pedagogical review, see, e.g., Ref.~\cite{Bambi:2017iyh}. More specifically, thermal photons from the accretion disk can inverse Compton scatter off free electrons in the corona. A fraction of the Comptonized photons can then illuminate the disk, generating the reflection spectrum.

In the rest-frame of the gas in the disk, the reflection spectrum is characterized by a soft excess below 2~keV, narrow fluorescent emission lines in the 1-10~keV band (including the iron K$\alpha$ line at 6-7~keV, depending on the ionization of the iron ions), and a Compton hump peaked at 20-30~keV~\cite{George:1991jj,Magdziarz:1995dc,Ross:2005dm,Garcia:2010iz}. The spectrum of the whole disk observed far from the source is the result of relativistic effects occurring in the strong gravity region around the black hole (Doppler boosting, gravitational redshift, and light bending)~\cite{Fabian:1989ej,Laor:1991nc}. The narrow fluorescent emission lines in the gas rest-frame are thus broadened and skewed for a distant observer. Since the reflection spectrum is mainly produced from the very inner part of the accretion disk, the analysis of relativistic reflection features in the spectra of accreting black holes can be used to study the accretion process around these compact objects, measure black hole spins, and possibly test fundamental physics~\cite{Brenneman:2006hw,Reynolds:2013qqa,Cao:2017kdq,Tripathi:2018lhx,DeRosa:2018aka,Tripathi:2020dni}.

In the past decade, there has been significant progress in the analysis of relativistic reflection spectra, thanks both to the development of more sophisticated theoretical models to calculate synthetic reflection spectra and the advent of new observational facilities more suitable for the study of these features. However, even the latest and most advanced theoretical models rely on a number of simplifications~\cite{Bambi:2020jpe}, so caution is necessary in the attempts to get precision measurements of the properties of accreting black holes from the analysis of their reflection features.

A number of studies have reported very steep reflection emissivity profiles in the inner part of accretion disks, both for stellar-mass black holes in X-ray binaries and supermassive black holes in active galactic nuclei (AGN); see, e.g., \cite{Wilkins:2011kt,Fabian:2012kv,Wilkins:2015nfa,Xu:2018lom,Zhang:2019ldz}. Steep emissivity profiles in the inner part of an accretion disk can be naturally produced by a compact source very close to the black hole~\cite{Martocchia96,Dauser:2013xv}. In such a case, the strong light bending near the black hole can drive the radiation emitted from the corona to illuminate better the inner part of the accretion disk closer to the event horizon.

The ionization of the disk plays an important role in the shape of the reflection spectrum. The ionization parameter, normally indicated with the letter $\xi$ and measured in units of erg~cm~s$^{-1}$, is defined as
\be\label{eq-xi}
\xi = \frac{4 \pi F_X (r)}{n_{\rm e} (r)} \, ,
\ee
where $F_X$ is the X-ray flux from the corona illuminating the disk and $n_{\rm e}$ is the disk electron density. In general, $\xi$, $F_X$, and $n_{\rm e}$ are all functions of the radial coordinate $r$. The radial profile of $F_X$ is determined by the coronal geometry. For $n_{\rm e}$, the radial profile depends on the disk properties.

Theoretical models for the analysis of relativistic reflection features in the spectra of accreting black holes normally assume a constant ionization parameter over all of the disk. From an observational point of view, for most data the fit simply does not require any ionization gradient. When the inner edge of the accretion disk is very close to the black hole and the corona is compact and low, the Comptonized photons are highly focused on a small portion of the very inner part of the accretion disk, which we can expect to be approximated well by a one-ionization region. For instance, the study in Ref.~\cite{Matzeu:2020ypn} finds a very high inner emissivity index when the data are fit with a model with a broken power-law emissivity profile and a constant disk ionization, but when they allow for a non-constant ionization profile the data do not require any ionization gradient. On the other hand, if the radial profile of $F_X$ is steep, even the profile of the ionization parameter $\xi$ should be steep for any reasonable disk density profile $n_{\rm e}$. This point was investigated in Ref.~\cite{Svoboda:2012cy} and then in Ref.~\cite{Kammoun:2019lpy}. The conclusion of both studies is that fitting the data with a theoretical model employing a constant ionization parameter may lead to overestimate the steepness of the inner emissivity profile, which, in turn, can lead to inaccurate black hole spin measurements; see also Ref.~\cite{Shreeram:2019ejg}.

The latest versions of the relativistic reflection models {\tt relxill}~\cite{Dauser:2013xv,Garcia:2013lxa}, {\tt kyn}~\cite{Dovciak:2004gq}, and {\tt reltrans}~\cite{Ingram:2019qlb} offer the option to have a non-trivial ionization profile in the disk. In the present paper, we implement a non-trivial ionization profile in our reflection model {\tt relxill\_nk}~\cite{Bambi17,Abdikamalov19,Abdikamalov:2020oci}, which is an extension of the {\tt relxill} package to non-Kerr spacetimes~\cite{Bambi:2015kza}. We then use this new version of {\tt relxill\_nk} to analyze a 2019 \textsl{NuSTAR} observation of the black hole binary EXO~1846--031, as previous analyses of these data had assumed a constant ionization parameter and found a very steep inner emissivity index. For this source, we find that the model with a non-trivial ionization gradient can provide a better fit, but the impact on the estimates of the black hole spin or of the deformation parameter are weak and, in the end, negligible.

Our manuscript is organized as follows. In Section~\ref{s-ion}, we describe the implementation of a radial ionization profile in {\tt relxill\_nk}. In Section~\ref{s-exo}, we present the spectral analysis of the 2019 \textsl{NuSTAR} observation of EXO~1846--031 with constant and non-constant disk ionization profiles. We discuss our results in Section~\ref{s-con}.


\section{Radial disk ionization profiles \label{s-ion}}

The {\tt relxill\_nk} package has so far assumed a constant ionization parameter across the entire disk. Since it also employs a constant electron density, from Eq.~(\ref{eq-xi}) it would follow that $F_X$ is constant too. The question is whether such an approximation can be used without introducing undesirable large systematic uncertainties in the estimate of the model parameters, and in particular on the black hole spin and the deformation parameter of the {\tt relxill\_nk} model. This issue can be particularly relevant when the fit finds a very high inner emissivity profile, which is also the case in which we can get more precise measurements of the black hole spin and the deformation parameter.

{\tt relxillion\_nk} is the new flavor of the {\tt relxill\_nk} package that offers the possibility of a non-constant radial ionization profile. We employ an empirical power-law form for the radial profile of the ionization parameter 
\be\label{eq-xi-r} 
\xi (r) = \xi_0 \left( \frac{R_{\rm in}}{r} \right)^{\alpha_\xi} \, ,
\ee
where $\xi_0$ is the value of the ionization parameter at the inner edge of the accretion disk, $R_{\rm in}$, and $\alpha_\xi$ is the extra parameter (ionization index) in {\tt relxillion\_nk} to describe the ionization gradient. The standard {\tt relxill\_nk} model with constant disk ionization is recovered for $\alpha_\xi = 0$, while for any positive value of $\alpha_\xi$ the value of the ionization parameter decreases as the radial coordinate $r$ increases.

The accretion disk is divided into 50 annuli and the value of the ionization parameter at the center of every annulus derived from Eq.~(\ref{eq-xi-r}). For every annulus, the reflection component is extracted from the {\tt xillver} table according to the ionization state of that region. The reflection spectrum from each annulus is then convolved to obtain the reflection spectrum detected by the distant observer and the spectra of all annuli are summed up together to obtain the total spectrum.

\begin{figure*}
    \includegraphics[width=0.99\textwidth]{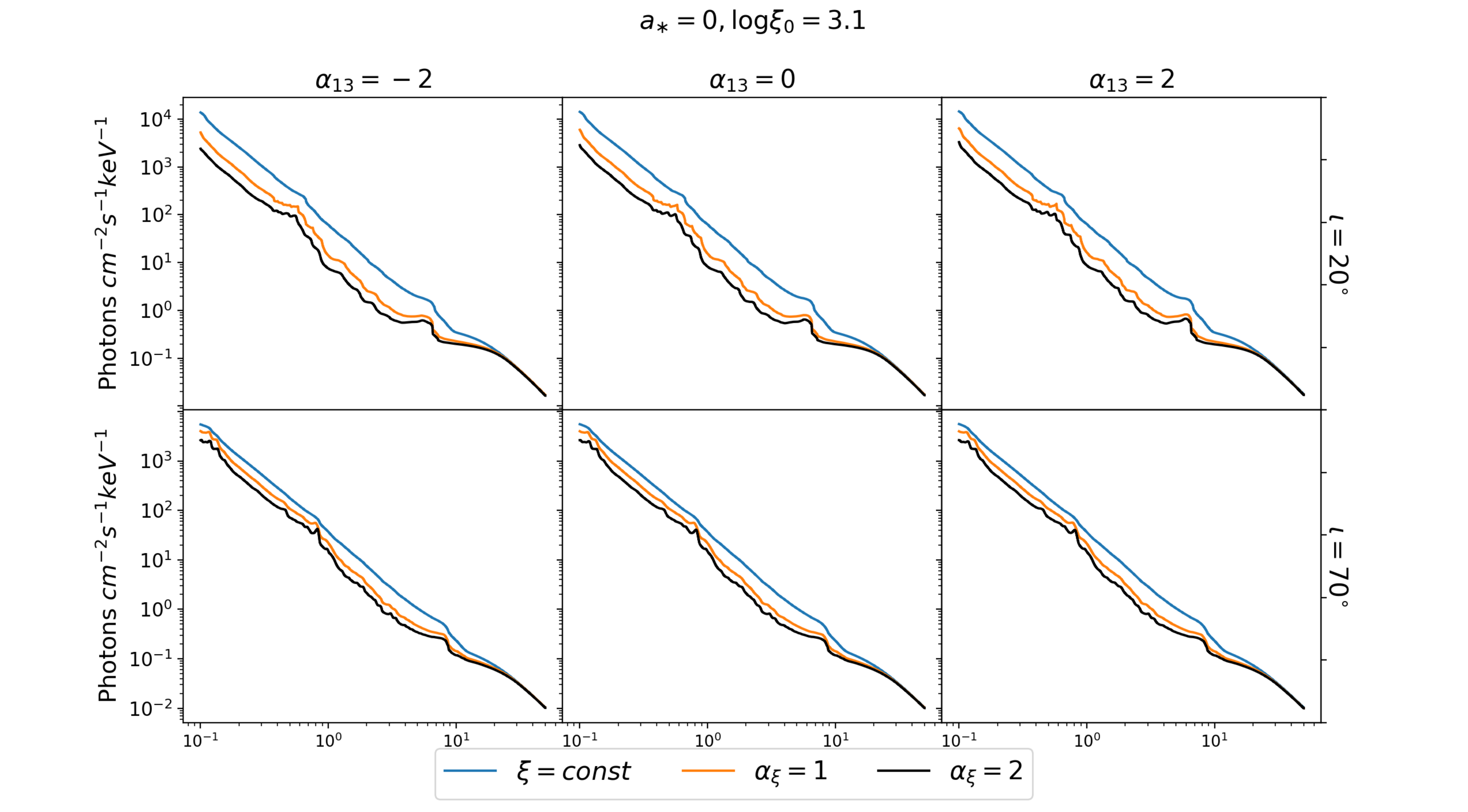}
    \caption{Synthetic relativistic reflection spectra in the Johannsen spacetime for the spin parameter $a_* = 0$, ionization parameter at the ISCO $\log\xi_0=3.1$, deformation parameter $\alpha_{13} = -2$, 0, and 2, and viewing angle $\iota = 20^\circ$ and $70^\circ$. The spectra with constant ionization parameter are in blue ($\alpha_\xi = 0$ and thus $\xi = \xi_0$), those with ionization index $\alpha_\xi=1$ are in orange, and the spectra with $\alpha_\xi=2$ are in black. \label{f-plot1}}
\end{figure*}

\begin{figure*}
    \includegraphics[width=0.99\textwidth]{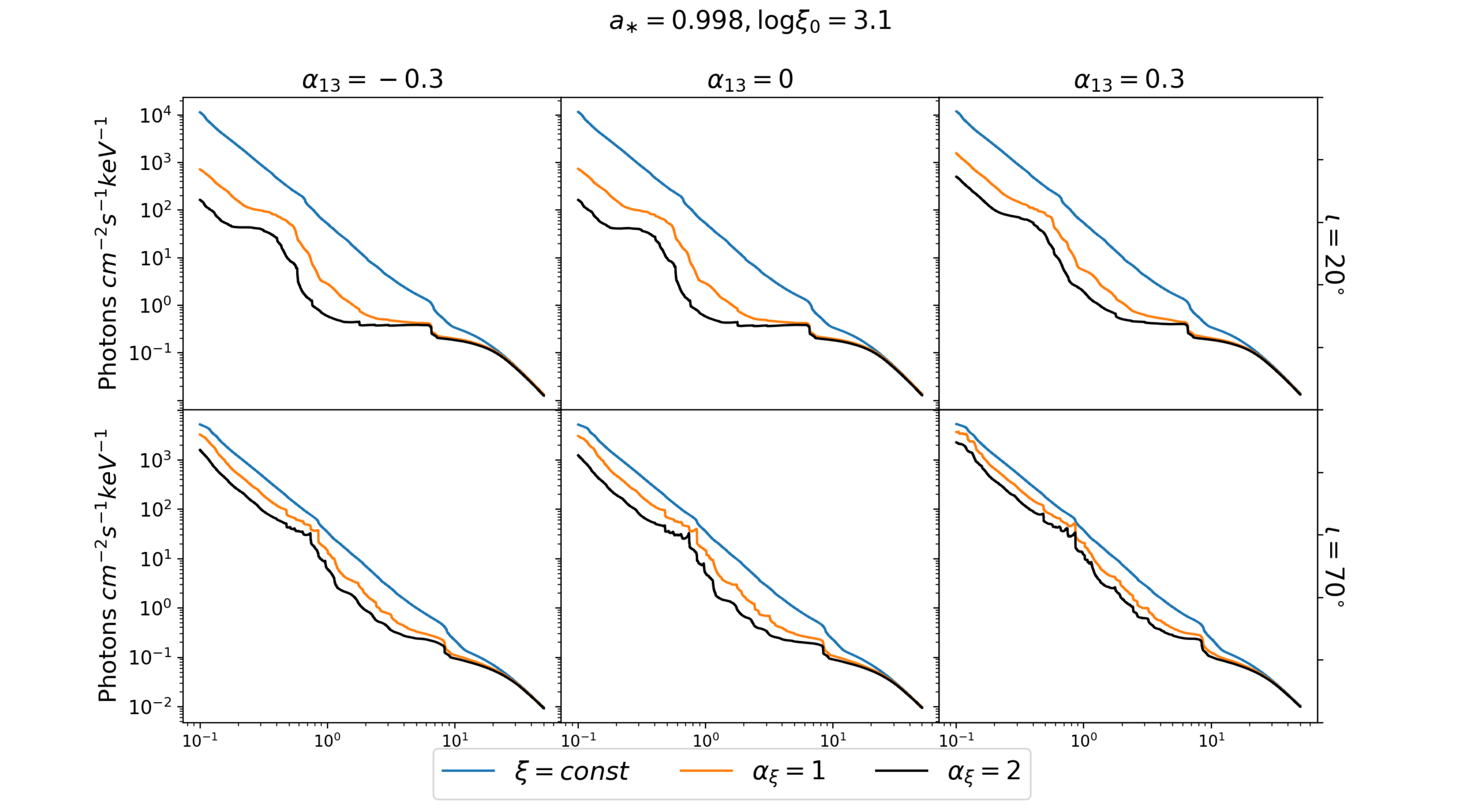}
    \caption{As in Fig.~\ref{f-plot1} for the spin parameter $a_* = 0.998$ and the Johannsen deformation parameter $\alpha_{13} = -0.3$, 0, and 0.3. \label{f-plot2}}
\end{figure*}

\begin{figure*}
    \includegraphics[width=0.99\textwidth]{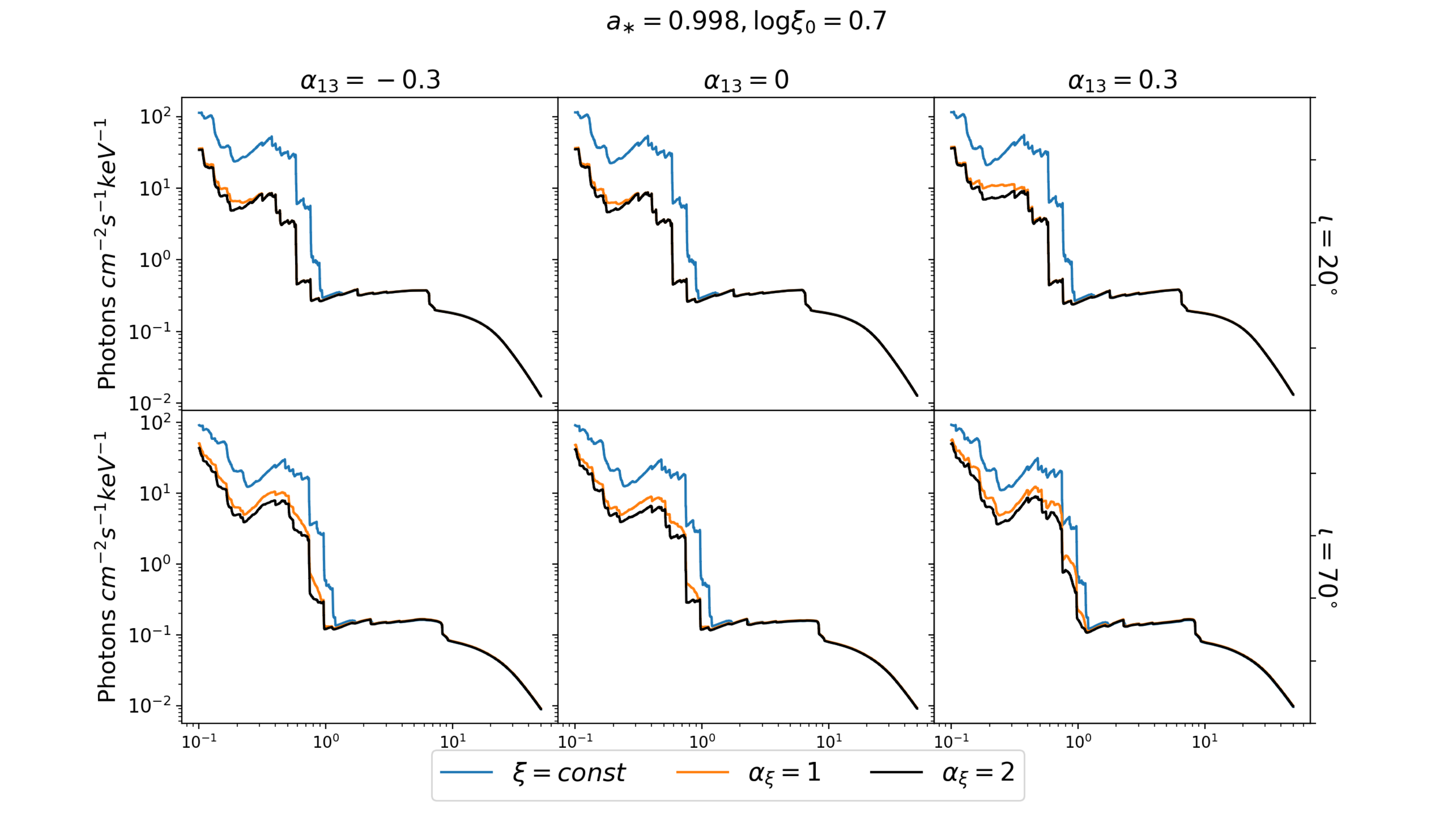}
    \caption{As in Fig.~\ref{f-plot2} for the ionization parameter at the ISCO $\log\xi_0 = 0.7$. \label{f-plot3}}
\end{figure*}

\begin{figure*}
    \includegraphics[width=0.99\textwidth]{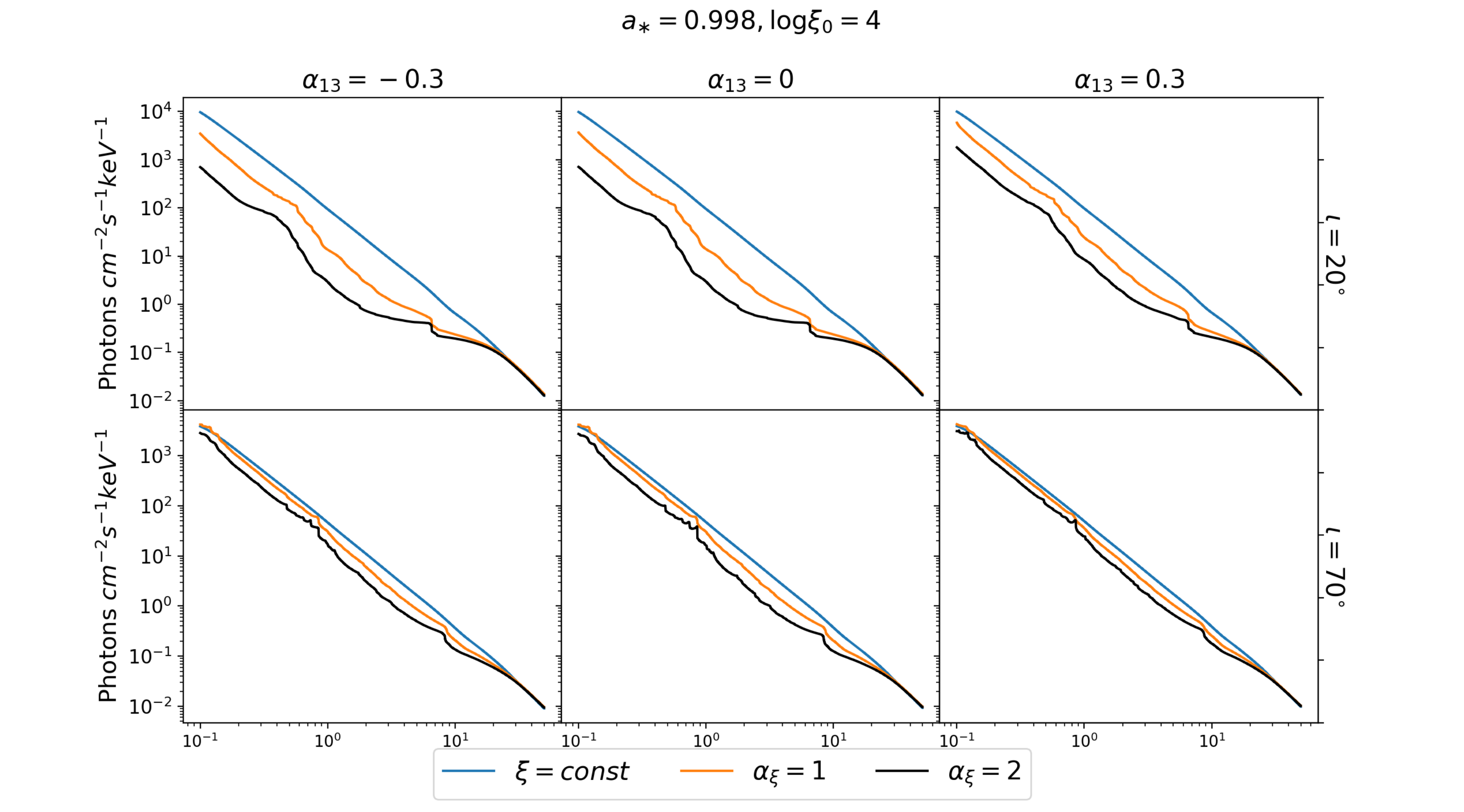}
    \caption{As in Fig.~\ref{f-plot2} for the ionization parameter at the ISCO $\log\xi_0 = 4$. \label{f-plot4}}
\end{figure*}

Figs.~\ref{f-plot1}-\ref{f-plot4} illustrate the impact of the ionization gradient of the accretion disk on the reflection spectrum of an accreting black hole. In all plots, we assume that the inner edge of the accretion disk is at the innermost stable circular orbit (ISCO), so $R_{\rm in} = R_{\rm ISCO}$ in Eq.~(\ref{eq-xi-r}). The constant ionization case ($\xi = \xi_0$, blue curves) is compared with the spectra calculated assuming the ionization index $\alpha_\xi=1$ (orange curves) and $\alpha_\xi=2$ (black curves) for various values of the black hole spin parameter $a_*$, the Johannsen deformation parameter $\alpha_{13}$\footnote{In this paper, we use the version of {\tt relxill\_nk} employing the Johannsen metric with non-vanishing deformation parameter $\alpha_{13}$~\cite{Johannsen13} as background metric.}, the viewing angle $\iota$, and the ionization parameter at the ISCO $\xi_0$.

The value of the ionization parameter $\xi$ determines the relative abundance of different ion species for every element and, in turn, the strength and the energy of the emission lines. For example, the iron K$\alpha$ line is at 6.4~keV in the case of neutral or weakly ionized iron and shifts up to 6.97~keV for H-like iron ions, while there is no line for fully ionized iron ions. In Fig.~\ref{f-plot3}, the ionization parameter at the inner edge of the disk is low, so we do not see difference in the three spectra above 1~keV because all models have a weakly ionized disk. In Fig.~\ref{f-plot4}, the ionization parameter at the inner edge of the disk is very high. In the model with constant ionization parameter, this means that the material in the whole disk is strongly ionized and the result is that its spectra do not show features (most ions are fully ionized). In the models with an ionization gradient, the spectra have features because the value of the ionization parameter decreases at larger radii. The case $\alpha_\xi = 2$ has more prominent features than the case $\alpha_\xi = 1$ just because the profile of the ionization parameter is steeper and we can see lines emitted from smaller radii.

In Figs.~\ref{f-plot1} and \ref{f-plot2}, the ionization parameter at the inner edge of the disk is moderately high, and therefore we have a situation between the cases in Fig.~\ref{f-plot3} and in Fig.~\ref{f-plot4}. For a low value of the black hole spin parameter (Fig.~\ref{f-plot1}), the inner edge of the disk is at a large radius, so the emission of the three models $\alpha_\xi = 0$, 1, and 2 differs at even larger radii, where relativistic effects are weaker and this limits the differences among the three models. For a high value of the black hole spin parameter (Fig.~\ref{f-plot2}), the relativistic effects in the very strong gravity region enhance the differences of the three models. The deformation parameter $\alpha_{13}$ has an impact qualitatively similar to the black hole spin: for a constant value of the black hole spin, the ISCO radius increases (decreases) if we increase (decrease) the value of $\alpha_{13}$, and this leads to smaller (larger) differences among the spectra with $\alpha_\xi = 0$, 1, and 2.


\section{The case of EXO~1846--031 \label{s-exo}}

EXO~1846--031 is a black hole candidate discovered in 1985 by the \textsl{EXOSAT} mission~\cite{Parmar85}. After spending more the 30~years in quiescence, it went through a new outburst in 2019~\cite{Negoro19}. \textsl{NuSTAR}~\cite{Harrison:2013md} observed EXO~1846--031 on 3 August 2019 (ObsID 90501334002) with a net exposure time of 22.2~ks. By analyzing its disk reflection features in the intermediate state of this outburst, it was found the system consists of a fast-rotating black hole and an accretion disk with high inclination angle~\cite{Draghis20}.

We reduce the same \textsl{NuSTAR} data as in~\cite{Draghis20} using the \texttt{nupipeline} and \texttt{nuproducts} routines in NuSTARDAS. The calibration database is CALDB 20200813. We extract the source spectra from circular regions with radii of 180$^{\prime\prime}$ on the FPMA and FPMB detectors. The background regions are of the same size but at the farthest diagonal place from the source regions to avoid the influence of the source photons. We group the spectra using \texttt{grppha} in order that each energy bin contains at least 30 counts. Since the new CALDB corrects the calibration in the 3-7~keV band, in the fitting we exclude the table \texttt{nuMLIv1.mod} used in~\cite{Draghis20}.

We perform the spectral fitting in XSPEC 12.10.1s~\cite{Arnaud96}. We first fit the spectra with an absorbed cut-off power-law, with a floating constant matching the slight discrepancy between FPMA and FPMB. The residuals below 4~keV indicate a disk thermal component, while the residuals around 6-7~keV and 10-30~keV indicate a reflection component. We thus use \texttt{diskbb}~\cite{Mitsuda84} and \texttt{relxill\_nk}~\cite{Bambi17,Abdikamalov19} to model these residuals, respectively. The final model set in XSPEC is 

\vspace{0.2cm}

\texttt{const$\times$tbabs$\times$(diskbb+relxill\_nk)} .

\vspace{0.2cm}

\noindent {\tt tbabs} describes the Galactic absorption and we leave the hydrogen column density $N_{\rm H}$ free in the fit~\cite{Wilms:2000ez}. {\tt diskbb} describes the thermal spectrum of a Newtonian disk\footnote{The component is weak, so it can already fit well with a simple Newtonian model and we do not need a more sophisticated one.} and the inner temperature of the disk, $T_{\rm in}$, is left free in the fit. For the reflection component, we use the flavor \texttt{relxillion\_nk} described in the previous section. We employ a broken power-law for the description of the emissivity profile, so we have the inner emissivity index ($q_{\rm in}$), the outer emissivity index ($q_{\rm out}$), and the breaking radius ($R_{\rm br}$). 
We assume that the inner radius of the disk is at the ISCO and the outer radius is fixed at 400 gravitational radii ($r_{\rm g}$). The spacetime metric is specified by the black hole spin $a_*$, which is allowed to vary between $-0.998$ and 0.998, and the Johannsen deformation parameter $\alpha_{13}$. The viewing angle $\iota$ is the angle between the line of sight and the axis of the black hole spin. The ionization of the disk is specified by $\xi_0$ and $\alpha_\xi$, as seen in the previous section. The iron abundance $A_{\rm Fe}$ is expressed in units of solar iron abundance. The model includes also the spectrum of the corona, so we have the photon index $\Gamma$, the high energy cut-off $E_{\rm cut}$, and the reflection fraction $R_{\rm f}$. Since EXO~1846--031 is a Galactic black hole, the redshift is set to 0. We fit the data with four models\footnote{We remind the reader that the case $\alpha_{13}=0$ corresponds to the Kerr solution, while deviations from the Kerr metric are present for a non-vanishing value of $\alpha_{13}$.}:

\vspace{0.2cm}

Model~1: $\alpha_\xi = 0$ and $\alpha_{13}=0$.

\vspace{0.2cm}

Model~2: $\alpha_\xi = 0$ and $\alpha_{13}$ free.

\vspace{0.2cm}

Model~3: $\alpha_\xi$ free and $\alpha_{13}=0$.

\vspace{0.2cm}

Model~4: $\alpha_\xi$ free and $\alpha_{13}$ free.

\vspace{0.2cm}

\begin{figure*}
\includegraphics[width=0.49\textwidth]{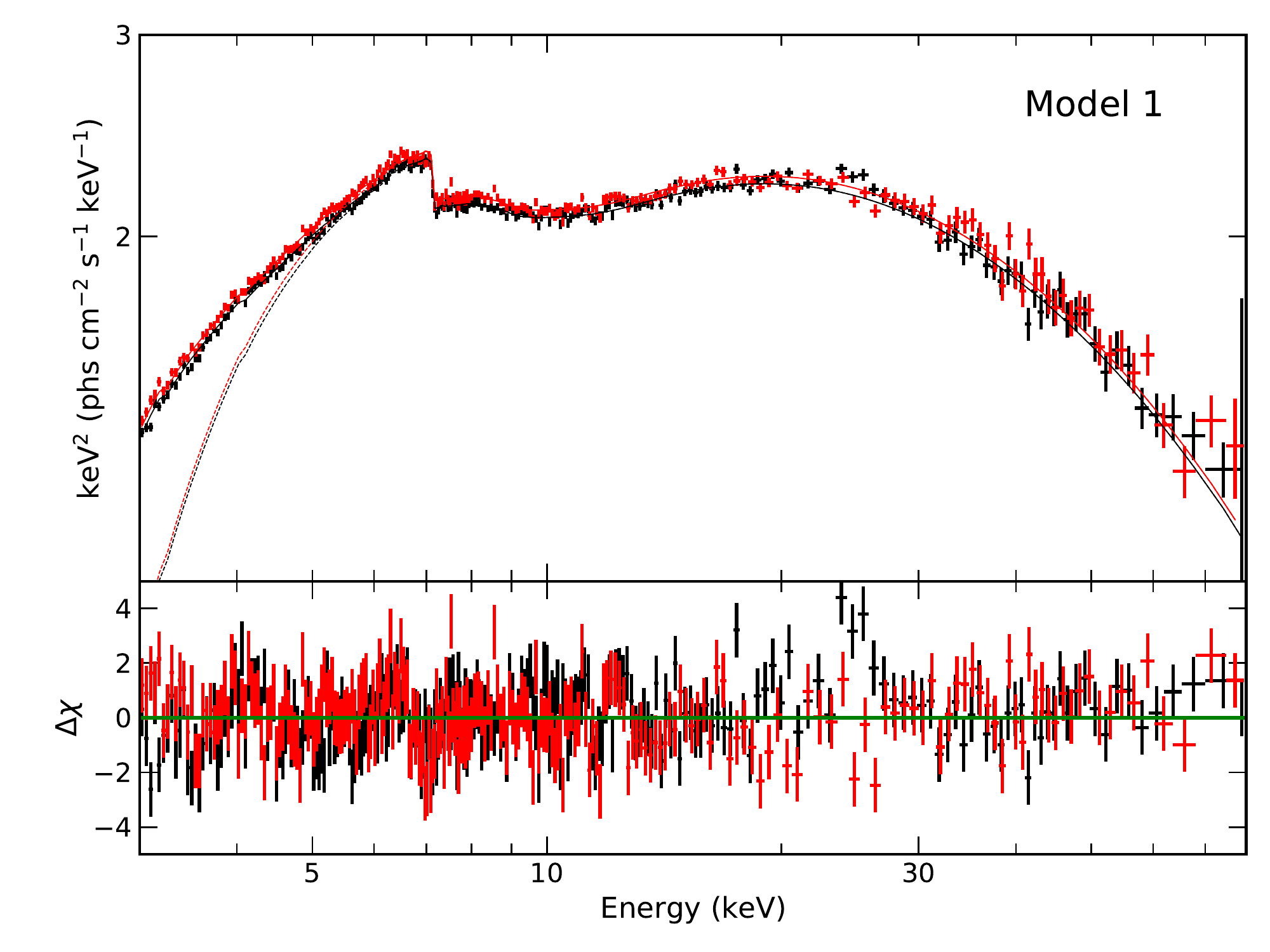}
\includegraphics[width=0.49\textwidth]{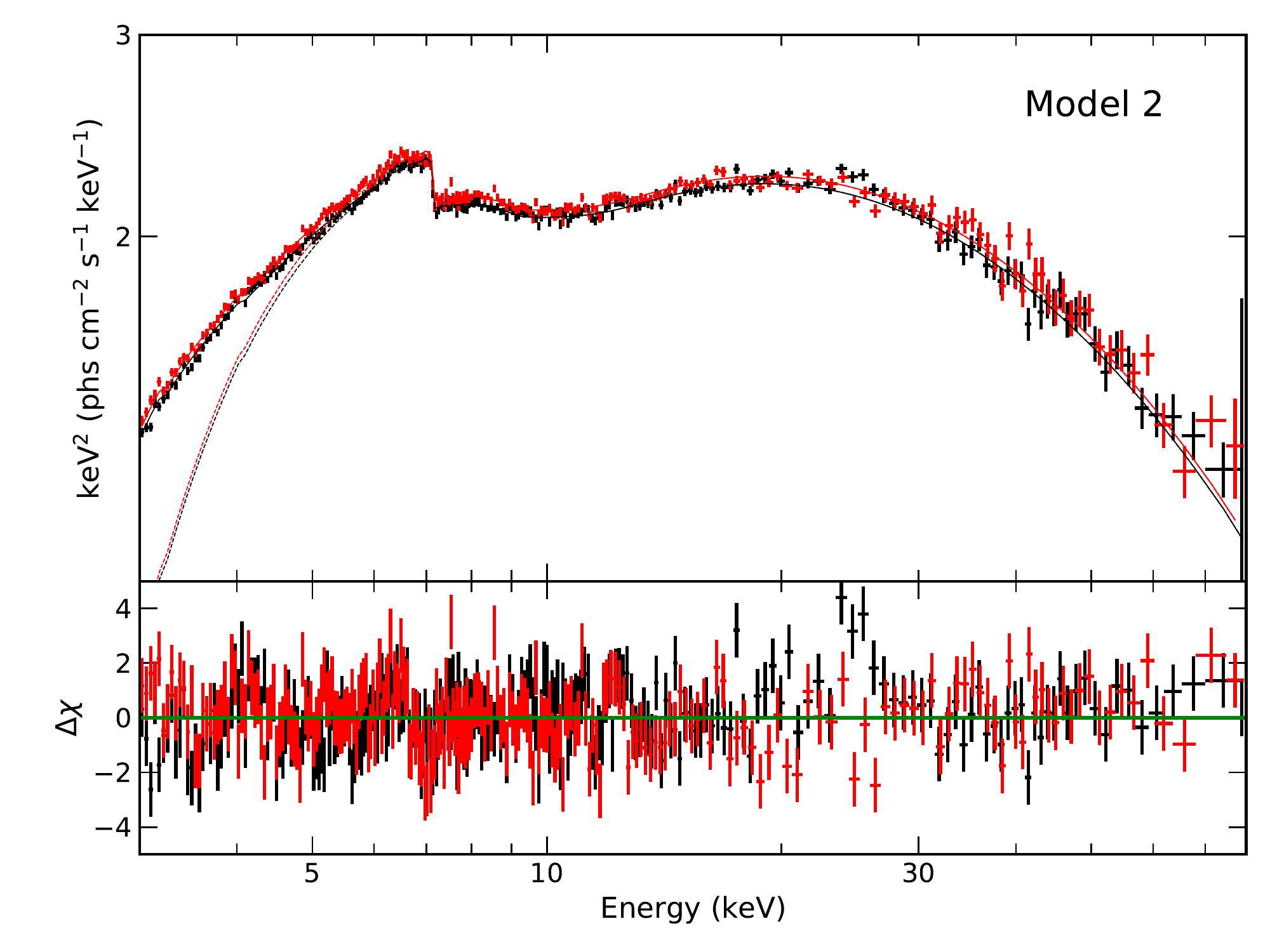} \\
\includegraphics[width=0.49\textwidth]{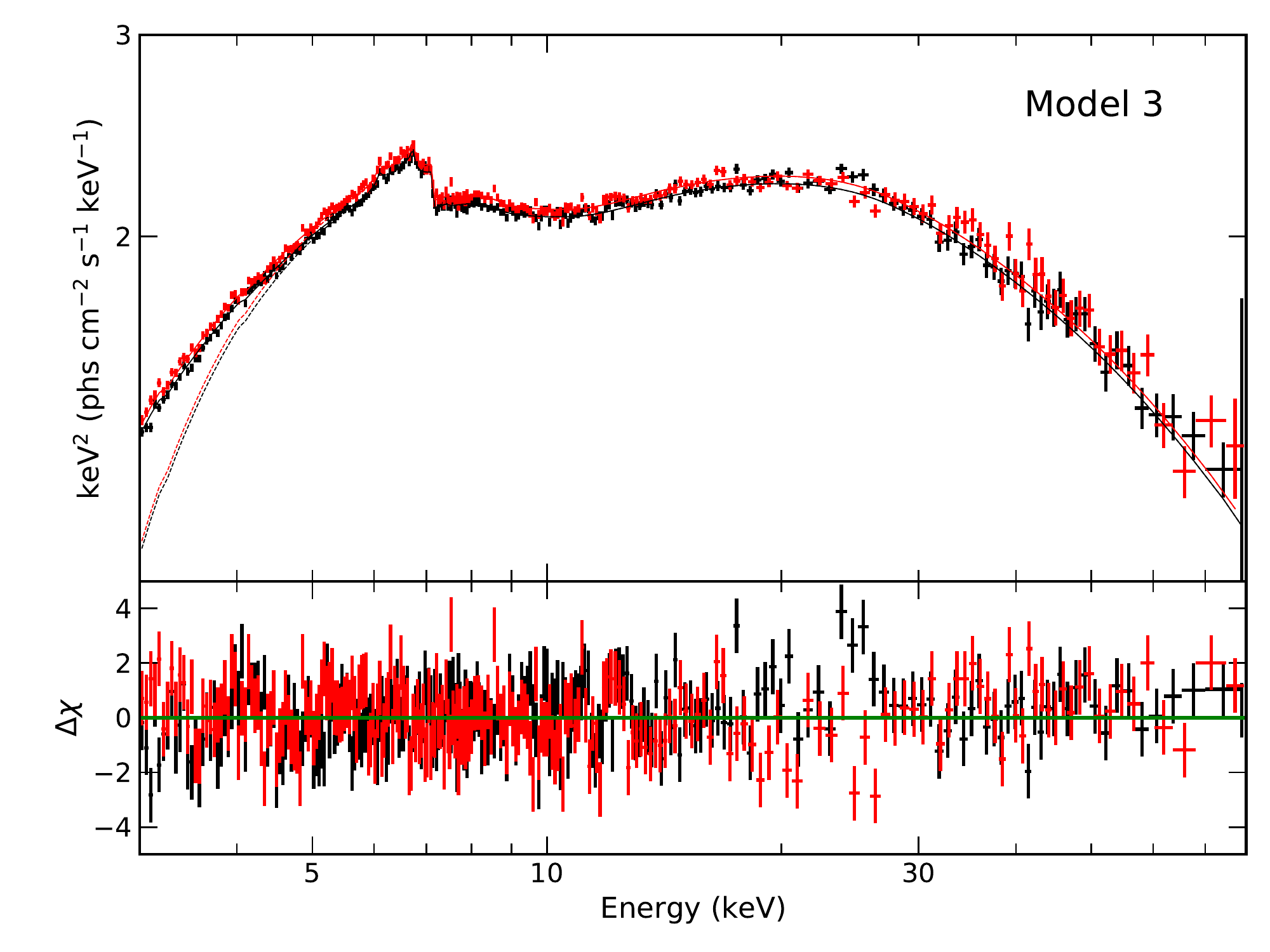}
\includegraphics[width=0.49\textwidth]{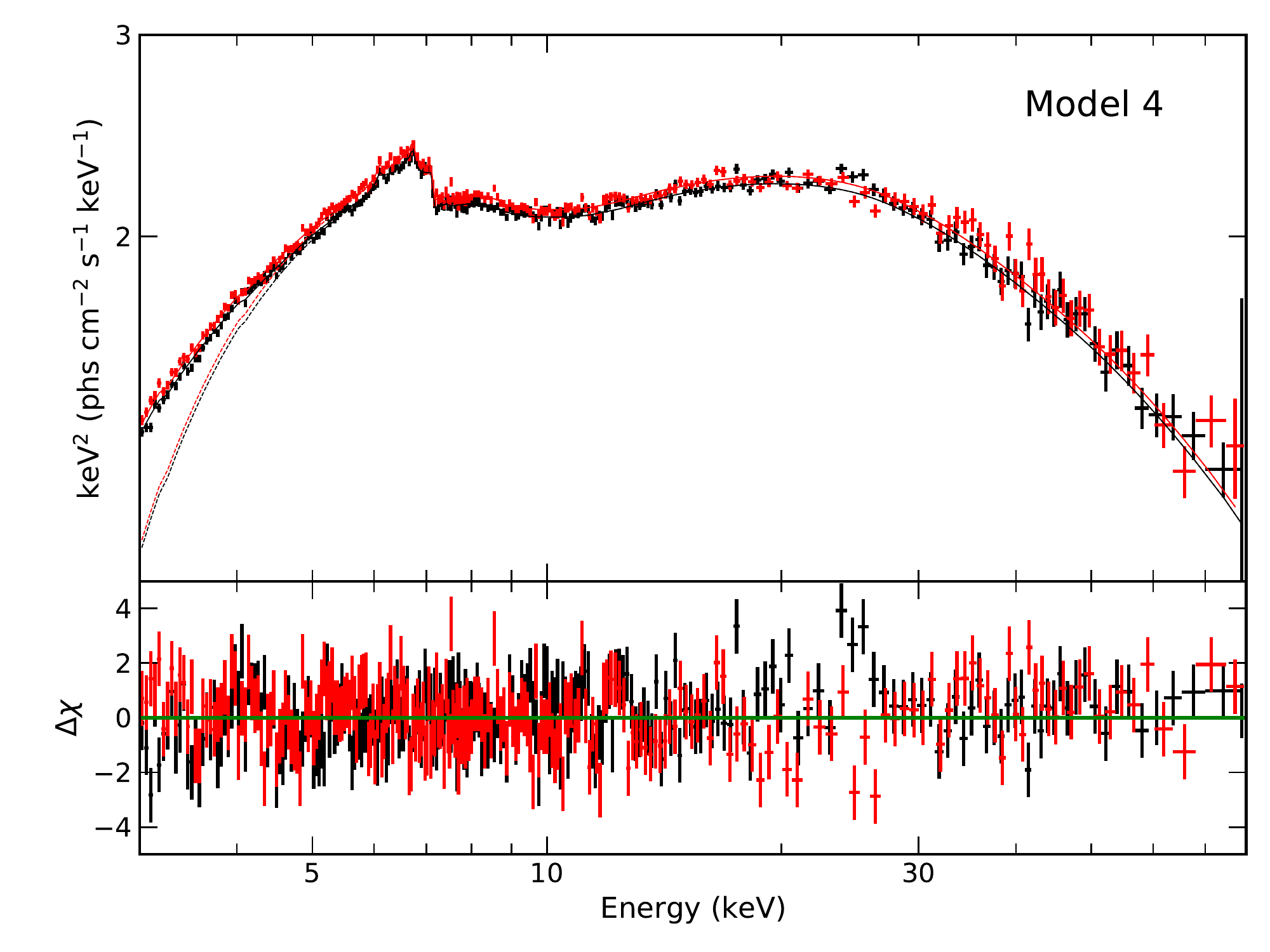}
\vspace{-0.2cm}
\caption{Best-fit models and residuals for Models~1-4. Black color for FPMA data and red color for FPMB data. In the best-fit model plots, solid curves for the total models and dotted curves for the reflection components (the thermal components are too weak and cannot be seen in the plots). \label{f-mr}}
\end{figure*}

Table~\ref{tab:fit} shows the best-fit values of the four fits. Best-fit models and ratio plots are in Fig.~\ref{f-mr}. Our results are discussed in the next section.

\begin{table*}
    \centering
    \renewcommand\arraystretch{1.3}{
    \begin{tabular}{lcccc}
        \hline\hline
        & Model~1 & Model~2 & Model~3 & Model~4 \\
        \hline
        \texttt{tbabs} &&&& \\
        $N_{\text{H}}$/$10^{22}$ cm$^{-2}$ & $10.1_{-0.4}^{+0.6}$ & $10.1_{-0.9}^{+0.6}$ & $8.1^{+0.9}_{-0.9}$ & $7.9^{+0.8}_{-1.0}$ \\
        \hline
        \texttt{diskbb} &&&& \\
        $kT_{\text{in}}$ [keV] & $0.458_{-0.011}^{+0.004}$ & $0.457_{-0.006}^{+0.009}$ & $0.41^{+0.03}_{-0.03}$ & $0.41^{+0.03}_{-0.04}$ \\
        norm [$10^{4}$] & $2.1_{-0.2}^{+0.2}$ & $2.1_{-0.9}^{+0.4}$ & $2.6^{+1.7}_{-1.0}$ & $2.3^{+1.9}_{-0.8}$ \\
        \hline
        \texttt{relxillion\_nk} &&&& \\
        $q_{\rm in}$ & $10.0_{-0.5}$ & $10_{-3}$ & $9.2_{-1.7}^{\rm +(P)}$ & $6.0^{+0.4}_{-0.8}$ \\
        $q_{\rm out}$ & $0.00^{+0.03}$ & $0.0^{+0.8}$ & $0.0^{+0.7}$ & $0.0^{+0.7}$ \\
        $R_{\text{br}}$ [$r_{\text{g}}$] & $6.5_{-0.1}^{+0.7}$ & $6.4_{-1.2}^{+0.2}$ & $7.1^{+4.7}_{-1.2}$ & $11^{+16}_{-3}$ \\
        $a_*$ & $0.9973_{-0.0009}^{\rm +(P)}$ & $0.9974_{-0.0004}^{\rm +(P)}$ & $0.980^{+0.009}_{-0.016}$ & $0.998_{-0.007}$ \\
        $\alpha_{13}$ & $0^*$ & $0.0_{-0.4}^{+0.1}$ & $0^*$ & $-0.3^{+0.3}_{-0.1}$ \\
        $\iota$ [deg] & $74.8^{+1.0}_{-0.1}$ & $74.8^{+0.6}_{-0.1}$ & $70^{+5}_{-2}$ & $71.0^{+6.5}_{-0.7}$ \\
        $\Gamma$ & $1.90_{-0.04}^{+0.03}$ & $1.90_{-0.02}^{+0.03}$ & $1.90^{+0.04}_{-0.07}$ & $1.87^{+0.06}_{-0.10}$ \\
        $\log\xi_0$ & $3.23_{-0.09}^{+0.07}$ & $3.26_{-0.05}^{+0.06}$ & $4.35^{+0.12}_{-0.14}$ & $4.48_{-0.09}^{\rm +(P)}$ \\
        $\alpha_\xi$ & $0^*$ & $0^*$ & $1.8^{+1.1}_{-0.8}$ & $1.2^{+0.4}_{-0.3}$ \\
        $A_{\text{Fe}}$ & $3.1_{-0.8}^{+0.2}$ & $3.1_{-0.2}^{+0.3}$ & $1.9^{+0.8}_{-0.5}$ & $2.4^{+2.1}_{-0.6}$ \\
        $E_{\text{cut}}$ [keV] & $80_{-7}^{+6}$ & $79_{-8}^{+6}$ & $126^{+17}_{-14}$ & $134^{+15}_{-30}$ \\
        $R_{\rm f}$ & $1.5_{-0.2}^{+0.4}$ & $1.6_{-0.2}^{+0.2}$ & $1.4^{+0.8}_{-0.5}$ & $1.7^{+1.1}_{-0.9}$ \\
        norm [$10^{-2}$] & $2.12_{-0.19}^{+0.05}$ & $2.10_{-0.10}^{+0.01}$ & $1.2_{-0.5}^{+0.5}$ & $1.0^{+0.6}_{-0.3}$ \\
        \hline
        $\chi^2/{\rm dof}$ & \hspace{0.2cm} $2727.27/2601$ \hspace{0.2cm} & \hspace{0.2cm} 2727.09/2600 \hspace{0.2cm} & \hspace{0.2cm} $2671.34/2600$ \hspace{0.2cm} & \hspace{0.2cm} $2670.48/2599$ \hspace{0.2cm} \\
        & $=1.04855$ & $=1.04888$ & $=1.02744$ & $=1.02750$ \\
        \hline\hline
    \end{tabular} }
\caption{Best-fit table for Models~1-4. The reported uncertainties correspond to the 90\% confidence level for one relevant parameter ($\Delta\chi^2 = 2.71$). $^*$ means the value is frozen in the fit. When there is no lower/upper uncertainty, the parameter is stuck at the lower/upper boundary of the range in which it is allowed to vary. (P) indicates that the 90\% confidence level region reaches the boundary. The ionization parameter $\xi_0$ is in units of erg~cm~s$^{-1}$. \label{tab:fit}}
\end{table*}

\begin{figure}
\includegraphics[width=0.47\textwidth]{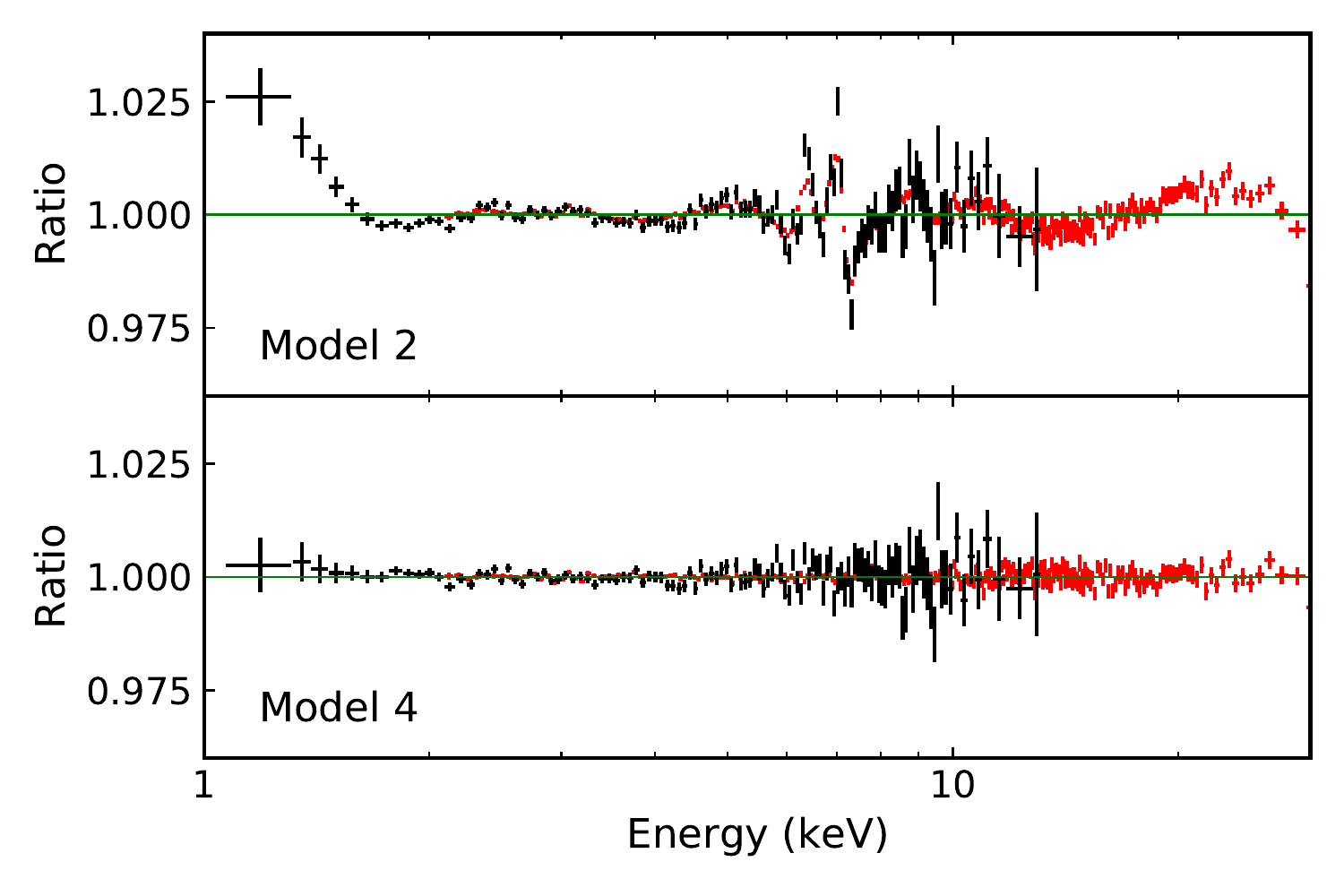}
\vspace{-0.4cm}
\caption{Data to the best-fit model ratio for Model~2 and Model~4 of the simulated 20~ks observation of EXO~1846--031 with X-IFU/\textsl{Athena} and LAD/\textsl{eXTP}. Black data for X-IFU/\textsl{Athena} and red data for LAD/\textsl{eXTP}. See the text for more details. \label{f-sim}}
\end{figure}

\section{Discussion and conclusions \label{s-con}}

The analysis of relativistic reflection features in black hole binaries and AGN shows that steep and very steep radial emissivity profiles are common in the very inner part of the accretion disk of black holes. While steep emissivity profiles can be naturally produced by compact coronae very close to the black hole, there is some debate whether current measurements of inner emissivity indices are overestimated. In particular, from Eq.~(\ref{eq-xi}) we should expect that even the radial profile of the ionization parameter $\xi$ is steep for any reasonable radial profile of the disk electron density $n_{\rm e}$, while most analyses assume a constant ionization parameter over all of the disk and several past attempts to fit the data with a non-trivial ionization profile found that an ionization gradient is not required by the data.

In the present work, we have presented an extension of our reflection model {\tt relxill\_nk} to include the possibility of a non-trivial radial disk ionization profile. In the new flavor, called {\tt relxillion\_nk}, the disk ionization profile is described by a power-law, so we have now the ionization parameter at the inner edge of the disk, $\xi_0$, and the ionization index $\alpha_\xi$ ($> 0$).

In the previous section, we have applied our new model to a 2019 \textsl{NuSTAR} observation of the black hole binary EXO~1846--031. First, we have fit the data assuming a constant ionization for all of the disk, $\alpha_\xi = 0$, and a broken power-law emissivity profile, either imposing the Kerr metric ($\alpha_{13} = 0$) and relaxing such an assumption ($\alpha_{13}$ free). In both cases, we find a very steep inner emissivity index, with $q_{\rm in}$ stuck to the maximum value allowed in our fit, and an almost vanishing outer emissivity index $q_{\rm out}$. We note that such an emissivity pattern could be generated by a number of coronal geometries, see Refs.~\cite{Wilkins:2015nfa,Miniutti:2003yd,Gonzalez:2017gzu}. The second model with $\alpha_{13}$ free does not improve the fit ($\Delta\chi^2 = 0.19$ with one less dof), so the data are consistent with the hypothesis that the black hole in EXO~1846--031 is a Kerr black hole as predicted by General Relativity.

The \textsl{NuSTAR} data are then fit with a free ionization index $\alpha_\xi$ and still assuming a broken power-law emissivity profile. As before, first we impose the Kerr metric  ($\alpha_{13} = 0$) and then we repeat the analysis with $\alpha_{13}$ free. The difference between the two fits is marginal ($\Delta\chi^2 = 0.86$) but the difference with the two fits with constant ionization parameter is not negligible ($\Delta\chi^2 \approx 56$). If we look at the ratio plots in Fig.~\ref{f-mr}, we see that the models with an ionization gradient can fit better the iron line. If we compare the estimates of the parameters inferred with $\alpha_\xi = 0$ and $\alpha_\xi$ free, Table~\ref{tab:fit}, we see that the ionization gradient has quite a modest impact on the measurements of the black hole spin parameter $a_*$ and on the deformation parameter $\alpha_{13}$. There is some difference in the estimate of $q_{\rm in}$, which is not stuck to the boundary any longer. The estimate of the hydrogen column density in {\tt tbabs} is a bit lower when $\alpha_\xi$ is free. The best-fit values of the iron abundance are lower with $\alpha_\xi$ free, but actually the uncertainties on the estimates are larger, so we cannot really say that the model with a radial disk ionization profile can find a lower iron abundance. The estimate of the high energy cut-off $E_{\rm cut}$ is higher with $\alpha_\xi$ free. In general, the uncertainties in the best-fit values are larger in the fits with $\alpha_\xi$ free than in those with $\alpha_\xi = 0$.

We note that in Ref.~\cite{Shreeram:2019ejg} the authors analyzed the 2012 \textsl{NuSTAR} observation of the black hole binary GRS~1915+105 and found that the ionization profile has an impact on the estimate of the black hole spin, in agreement with the claim in Ref.~\cite{Kammoun:2019lpy}, where the authors only worked on simulations and limited their study to the 1-10~keV band. The conclusion of those authors is in apparent disagreement with our results, where we do not find any bias on the black hole spin measurement (and the deformation parameter). However, the shape of relativistic reflection spectra is determined by many parameters and it is possible that in certain regions of the parameter space the ionization gradient can be ignored and in other regions it cannot, as well as that in some regions the correct ionization gradient has an impact on the estimate of some model parameters and in other regions it does not. Note also that both those papers used a lamppost model, while here we use a broken power-law for the emissivity profile.

To explore the impact of a disk ionization gradient on higher quality data with future X-ray missions, we simulate a 20~ks observation of EXO~1846--031 with X-IFU/\textsl{Athena}~\cite{Nandra:2013jka} and LAD/\textsl{eXTP}~\cite{Zhang:2016ach}. As input values, we use the best-fit values of Model~3. We then fit the simulated data with Model~2 and Model~4. This time the model with constant ionization parameter cannot fit the data well. The data to the best-fit model ratios of the two fits are shown in Fig.~\ref{f-sim}, where we clearly see that Model~2 cannot fit well the iron line region, the 1-2~keV band, and the Compton hump. The reduced $\chi^2$ is 1.20 for Model~2 and 1.01 for Model~4. The large effective areas of X-IFU/\textsl{Athena} and LAD/\textsl{eXTP} lead to very precise measurements of the model parameters, so we can see the impact of the ionization gradient. In particular, the estimate of the Johannsen deformation parameter is $\alpha_{13} = 0.2057_{-0.0019}^{+0.004}$ with Model~2, and thus we do not recover Kerr even if the simulation assumes the Kerr spacetime, and $\alpha_{13} = 0.0025_{-0.004}^{+0.0008}$ with Model~4. Our simulations thus suggest that tests of the Kerr metric with the next generation of X-ray mission require reflection models that include the ionization profile of the disk; models with a constant ionization profile may incorrectly infer non-vanishing deformation parameters even if the source is a Kerr black hole.


\vspace{0.5cm}

{\bf Acknowledgments --}
This work was supported by the Innovation Program of the Shanghai Municipal Education Commission, Grant No.~2019-01-07-00-07-E00035, the National Natural Science Foundation of China (NSFC), Grant No.~11973019, and Fudan University, Grant No.~JIH1512604.
D.A. is supported through the Teach@T{\"u}bingen Fellowship.
Y.Z. acknowledges the support from China Scholarship Council (CSC 201906100030).
C.B. and H.L. are members of the International Team~458 at the International Space Science Institute (ISSI), Bern, Switzerland, and acknowledge support from ISSI during the meetings in Bern.


\end{document}